\documentclass{article}

\usepackage{arxiv}

\usepackage[utf8]{inputenc} 
\usepackage[T1]{fontenc}    
\usepackage{hyperref}       
\usepackage{url}            
\usepackage{booktabs}       
\usepackage{amsfonts}       
\usepackage{nicefrac}       
\usepackage{microtype}      
\usepackage{lipsum}		
\usepackage{graphicx}
\usepackage{natbib}
\usepackage{doi}
\usepackage{amsmath,amssymb,amsfonts}
\usepackage{multirow}
\usepackage{tablefootnote}
\usepackage{algorithmic}
\usepackage{graphicx}
\usepackage{verbatim}
\usepackage{textcomp}
\usepackage{booktabs}
\usepackage[normalem]{ulem}
\newcommand{\ra}[1]{\renewcommand{\arraystretch}{#1}}

\title{EEG-Based Epileptic Seizure Prediction Using Temporal Multi-Channel Transformers}

\author{ \href{https://orcid.org/0000-0002-5323-9299}{\includegraphics[scale=0.06]{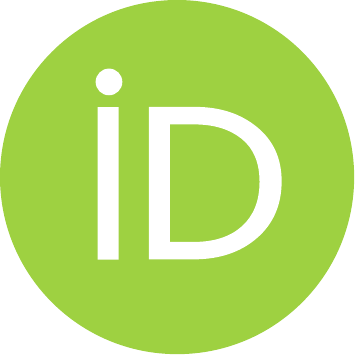}\hspace{1mm}~Ricardo V. Godoy} \\
	  São Carlos School of Engineering\\
	University of São Paulo\\
	São Carlos, Brazil \\
	\texttt{ricardo.godoy@alumni.usp.br} \\
	\And
	{\includegraphics[scale=0.0]{orcid.pdf}\hspace{1mm}Tharik J. S. Reis} \\
	  São Carlos School of Engineering\\
	University of São Paulo\\
	São Carlos, Brazil \\
	\And
	{\includegraphics[scale=0.0]{orcid.pdf}\hspace{1mm}Paulo H. Polegato} \\
	  São Carlos School of Engineering\\
	University of São Paulo\\
	São Carlos, Brazil \\	\And
        \href{https://orcid.org/0000-0002-0403-991X}
	{\includegraphics[scale=0.06]{orcid.pdf}\hspace{1mm}Gustavo J. G. Lahr} \\
	  Human-Robot Interfaces and Physical Interaction Lab\\
	Istituto Italiano di Tecnologia\\
	Genoa, Italy \\	\And
	\href{https://orcid.org/0000-0002-5570-9055}
	{\includegraphics[scale=0.06]{orcid.pdf}\hspace{1mm}Ricardo L. Saute} \\
	  Medical School of Ribeirão Preto\\
	University of São Paulo\\
	Ribeirão Preto, Brazil \\	\And
	{\includegraphics[scale=0.0]{orcid.pdf}\hspace{1mm}Frederico N. Nakano} \\
	  Medical School of Ribeirão Preto\\
	University of São Paulo\\
	Ribeirão Preto, Brazil \\	\And
	\href{https://orcid.org/0000-0002-1069-7804}
	{\includegraphics[scale=0.06]{orcid.pdf}\hspace{1mm}Helio R. Machado} \\
	  Medical School of Ribeirão Preto\\
	University of São Paulo\\
	Ribeirão Preto, Brazil \\	\And
	{\includegraphics[scale=0.0]{orcid.pdf}\hspace{1mm}Americo C. Sakamoto} \\
	  Medical School of Ribeirão Preto\\
	University of São Paulo\\
	Ribeirão Preto, Brazil \\	\And
	\href{https://orcid.org/0000-0002-7508-5817}
	{\includegraphics[scale=0.06]{orcid.pdf}\hspace{1mm}Marcelo Becker} \\
	  São Carlos School of Engineering\\
	University of São Paulo\\
	São Carlos, Brazil \\	\And
	\href{https://orcid.org/0000-0003-0898-1379}
	{\includegraphics[scale=0.06]{orcid.pdf}\hspace{1mm}Glauco A. P. Caurin} \\
	  São Carlos School of Engineering\\
	University of São Paulo\\
	São Carlos, Brazil \\
}

\date{April 19, 2022}

\renewcommand{\headeright}{ }
\renewcommand{\undertitle}{ }
\renewcommand{\shorttitle}{EEG-Based Epileptic Seizure Prediction Using Temporal Multi-Channel Transformers}

\hypersetup{
pdftitle={EEG Based Epileptic Seizure Prediction Using Temporal Multi-Channel Transformers},
pdfsubject={Deep learning, electroencephalography, epilepsy, seizure prediction, Transformers},
pdfauthor={Ricardo V. Godoy, Tharik J. S. Reis, Paulo H. Polegato, Gustavo J. G. Lahr, Ricardo L. Saute, Frederico N. Nakano, Helio R. Machado, Americo C. Sakamoto, Marcelo Becker, and Glauco A. P. Caurin},
pdfkeywords={Deep learning, electroencephalography, epilepsy, seizure prediction, Transformers},
}

\begin{document}
\maketitle

\begin{abstract}
Epilepsy is one of the most common neurological diseases, characterized by transient and unprovoked events called epileptic seizures. Electroencephalogram (EEG) is an auxiliary method used to perform both the diagnosis and the monitoring of epilepsy. Given the unexpected nature of an epileptic seizure, its prediction would improve patient care, optimizing the quality of life and the treatment of epilepsy. Predicting an epileptic seizure implies the identification of two distinct states of EEG in a patient with epilepsy: the preictal and the interictal. In this paper, we developed two deep learning models called Temporal Multi-Channel Transformer (TMC-T) and Vision Transformer (TMC-ViT), adaptations of Transformer-based architectures for multi-channel temporal signals. Moreover, we accessed the impact of choosing different preictal duration, since its length is not a consensus among experts, and also evaluated how the sample size benefits each model. Our models are compared with fully connected, convolutional, and recurrent networks. The algorithms were patient-specific trained and evaluated on raw EEG signals from the CHB-MIT database. Experimental results and statistical validation demonstrated that our TMC-ViT model surpassed the CNN architecture, state-of-the-art in seizure prediction.
\end{abstract}


\section{Introduction} \label{sec:introduction}

Epilepsy is a neurological disease characterized by the occurrence of transient and unprovoked events called epileptic seizures, resulting from changes in brain electrical activity \cite{Fisher2014b}. It is estimated that 1\% of the human population is affected by this disorder \cite{Weinstein2016}. Epilepsy is associated with stigma and adverse effects that, in addition to the psychological ones, extend to the social, cognitive, and economic fields \cite{Fisher2014b}. One-third of epilepsy patients cannot have their seizures controlled with two or more adequate trials of antiseizure medications, fulfilling the diagnosis of drug-resistant epilepsy \cite{Kwan2009}. Some patients are subjected to a neurological evaluation to assess their viability of undergoing non-pharmacological alternative treatment, such as neurosurgery, ketogenic diet, or responsive neurostimulation. Despite these possibilities, a significant number of patients persist with seizure recurrence. These patients suffer the most impact on quality of life and mortality, including sudden unexpected death in epilepsy,  which can occur during or after seizures \cite{459}. Furthermore, the unpredictable nature of an epileptic seizure, which ranges from minor distractions to violent convulsions \cite{Acharya2013}, has negative psychological impacts on the patient's life. Therefore, the possibility of predicting seizures could provide these patients a critical tool to avoid being caught off guard by seizures, to feel they have control over their bodies, and to have the opportunity to go to a safe place, get help, or even intervene through medication or neurostimulation to prevent the seizure.

\begin{figure}[!t]
\centerline{\includegraphics[width=0.80\columnwidth]{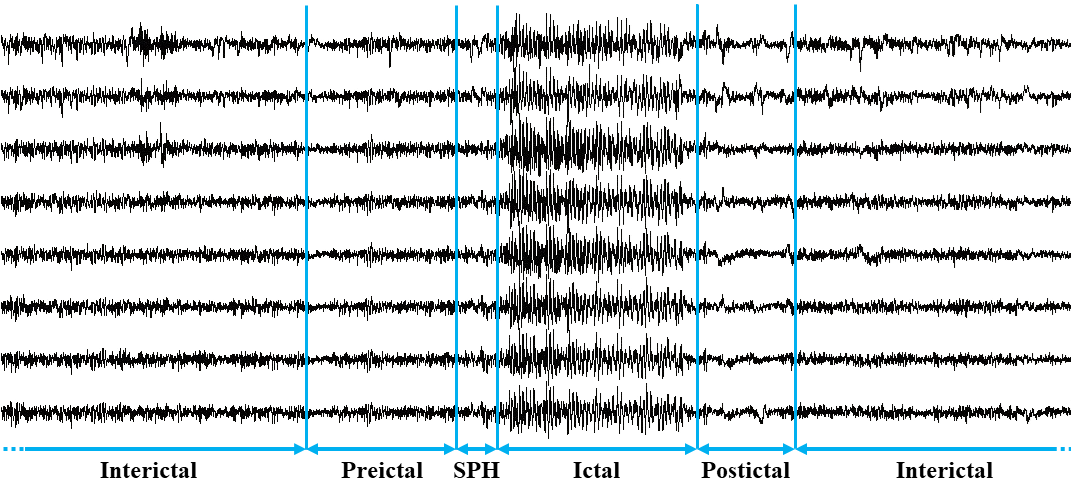}}
\caption{EEG signals during an absence seizure. EEG signals from patients with epilepsy are classified into four distinct states of brain activity: the preictal state, defined as a period before the onset of a seizure; the ictal state, which occurs during a seizure; the postictal state, which occurs after the seizure ends; and the interictal period, defined as the extent of time between seizures and is not one of the aforementioned states \cite{Chiang2011}. 
We defined a 5-minutes seizure prediction horizon (SPH) to guarantee response time for the medical staff and avoid the presence of ictal activity in the preictal period.}
\label{fig:eeg_state}
\end{figure}


One of the auxiliary methods to perform both the clinical diagnosis of epilepsy and the monitoring of patients is the electroencephalogram (EEG). Shown in Fig. \ref{fig:eeg_state}, EEG signals are recordings of electrical brain activity obtained by placing non-invasive electrodes on the patient's scalp.
The analysis of EEG signals is usually performed by a professional via visual inspection, and this is a costly process of time and human resources. Moreover, a visual inspection may detect and characterize the seizures, but it cannot identify the preictal state by its features without first detecting the subsequent seizure. The autonomous analysis of the signals obtained through EEG would lead to a more efficient process.

Artificial intelligence (AI), specifically machine learning (ML), has great potential to accomplish this task. Over the last few years, several studies have used ML-based methods to predict epileptic seizures. Among these methods, deep learning (DL) approaches have been gaining notoriety in the last 5 to 6 years \cite{Rasheed2021} due to their satisfactory results in addressing this issue. DL techniques identify and incorporate patterns from processed or raw data, resulting in an increasingly complex and robust system. Furthermore, EEG data are time-series signals, i.e., sequentially measured over time, which can lead to the occurrence of temporal dependencies. Fortunately, DL proves to be a tool with significant potential for analyzing this type of signal, enabling the learning of long-term characteristics and representing the state-of-the-art for various tasks of time series classification \cite{IsmailFawaz2019}.

Among the DL methods, the most used are convolutional neural networks (CNN) and recurrent neural networks (RNN). These methods will automatically extract and learn the relevant features from raw data, generating the desired output. Transformer architectures \cite{Vaswani2017}, developed to perform sequential data processing and solve problems such as the CNN's long training time and the RNN's inherent impossibility of parallelization, currently represent the state-of-the-art in several natural language processing (NLP) tasks. However, although multi-channel EEG signals are also sequential data, the Transformer architecture has not been used to process these signals.

This paper develops and presents an adaptation of the Transformer and Vision Transformer (ViT) for analyzing EEG signals to predict epileptic seizures. These architectures were named Temporal Multi-Channel Transformer (TMC-T) and Temporal Multi-Channel Vision Transformer (TMC-ViT), and they have already achieved outstanding results when using electromyography signals as input for both regression~\cite{godoy2022access,godoy2022tnsre} and classification~\cite{godoy2022ral} tasks. To evaluate the performance of these models, we compare results with three other classic DL techniques: multilayer perceptron (MLP), CNN, and a CNN concatenated to a bidirectional long short-term memory (Bi-LSTM). 

The contributions of this work are twofold: 1) Regarding the preictal definition, we conducted computational tests with different size definitions for this period since there is no consensus in the literature. The results obtained here can be used by future works to choose the preictal duration; 2) The adaptation of a Transformer and a ViT to use temporal data with multiple channels is a substantial contribution of this paper. The algorithms developed here can be used for future works involving time series and large amounts of data in areas not restricted to NLP or single-channel sequential data. 

The remaining sections are organized as follows. Section \ref{sec:related_work} presents an overview of the related work on seizure prediction using DL techniques and the Transformer architectures. Section \ref{sec:transformer_architectures} of the Transformers and ViT architectures. Section \ref{sec:models} presents an explanation concerning the models here implemented. Section \ref{sec:experiments} specifies the dataset employed for training and testing, the preprocessing steps, and how the models were trained and evaluated. The results are presented in Section \ref{sec:results}, whereas Section \ref{sec:conclusion} concludes the paper and discusses some potential future directions. 

\section{Related Work} \label{sec:related_work}

Several studies have used ML techniques to build patient-specific algorithms to classify EEG signals of patients with epilepsy and predict seizures. The most used dataset in the literature for this purpose is the CHB-MIT Scalp EEG Database \cite{Goldberger2000}, which contains EEG signals of several patients. 


Daoud and Bayoumi \cite{Daoud2019} employed a CNN concatenated to a Bi-LSTM and raw EEG signals as input for seizure prediction within 1 hour in advance. Wavelet (WL) packet decomposition and common spatial pattern (CSP), were designed to extract the distinguishing features in both the time domain and the frequency domain, and a CNN was employed by \cite{Zhang2020} for the task of seizure prediction. A hardware-friendly network called Binary Single-dimensional Convolutional Neural Network (BSDCNN) was implemented in \cite{Zhao2020}, using 1D convolutional kernels to improve prediction performance. A CNN + Extreme Learning Machine (ELM) scheme was proposed by \cite{Qin2020}, achieving good results using short-time Fourier transform (STFT) from raw data as input. A dual self-attention residual network (RDANet) was proposed by \cite{Yang2021}, where they converted raw EEG into spectrograms that represent time-frequency characteristics using STFT.



Despite clinical evidence that the preictal state exists \cite{Federico2005}, its approximate time period is unknown, and it may also be unique to each patient. Some papers \cite{Direito2017, Teixeira2014}, based on the European Epilepsy Database\footnote{The European Epilepsy Database requires a paid license to access it.} \cite{Ihle2012}, approached this issue by testing different times for the preictal state. These works tested their models for 10, 20, 30, and 40 min periods. Direito et al. \cite{Direito2017} used an SMV classifier and obtained an average preictal period of 28 minutes. Teixeira et al. \cite{Teixeira2014} employed SVM, MLP, and Radial Basis Functions Networks models, thus obtaining an average preictal period of 30 minutes.
However, for the CHB-MIT database, the largest open-access database and used in this paper, there is a deficiency in the justification for the definition of the preictal period: several authors divide it into 30 minutes \cite{Zhang2020, Zhao2020, Yang2021}, while others choose 60 minutes \cite{Qin2020, Daoud2019} before an epileptic seizure. The majority of researchers argue that there is no consensus among specialists in the medical field regarding the delimitation of this period so that the choice is made arbitrarily. It is noteworthy that, in some works, extracted features were used instead of raw EEG data. However, employing raw data may present advantages, such as the minimal preprocessing required, which can optimize the solutions for online applications, and the minimal need for prior knowledge of the data, in contrast to when feature engineering is employed.

Concerning the architectures used, a prevalence of CNN and RNN, specifically, the LSTM, is noticed. The Transformer architecture \cite{Vaswani2017}, mainly employed in NLP, represents the state-of-the-art in several language tasks, including translation \cite{Ott2018} and auto-regressive word generation \cite{Brown2020}. Recent research has developed Transformer architecture models for use in other tasks, such as the ViT for applications in the field of computer vision \cite{Dosovitskiy2020}. Regarding EEG signals, Krishna et al. \cite{Krishna2019} proposed an automatic speech recognition model based on Transformer using EEG features as input. In epilepsy, good results were obtained using a Transformer model to predict the response to different antiseizure medications in individuals with newly diagnosed epilepsy \cite{Choong2020}. These recent advances open up a new range of application areas, justifying its implementation to temporal signals and prediction of epileptic seizures, an area still unexplored by Transformers.

\section{Transformer Architectures} \label{sec:transformer_architectures}

As Transformers-based models for epileptic seizure prediction constitute a significant contribution of this work, this section aims to present the fundamental concepts of this architecture. Section \ref{sec:models} provides a thorough description of how these models were implemented and used in this work.

\begin{figure}[!t]
\centerline{\includegraphics[width=0.40\columnwidth]{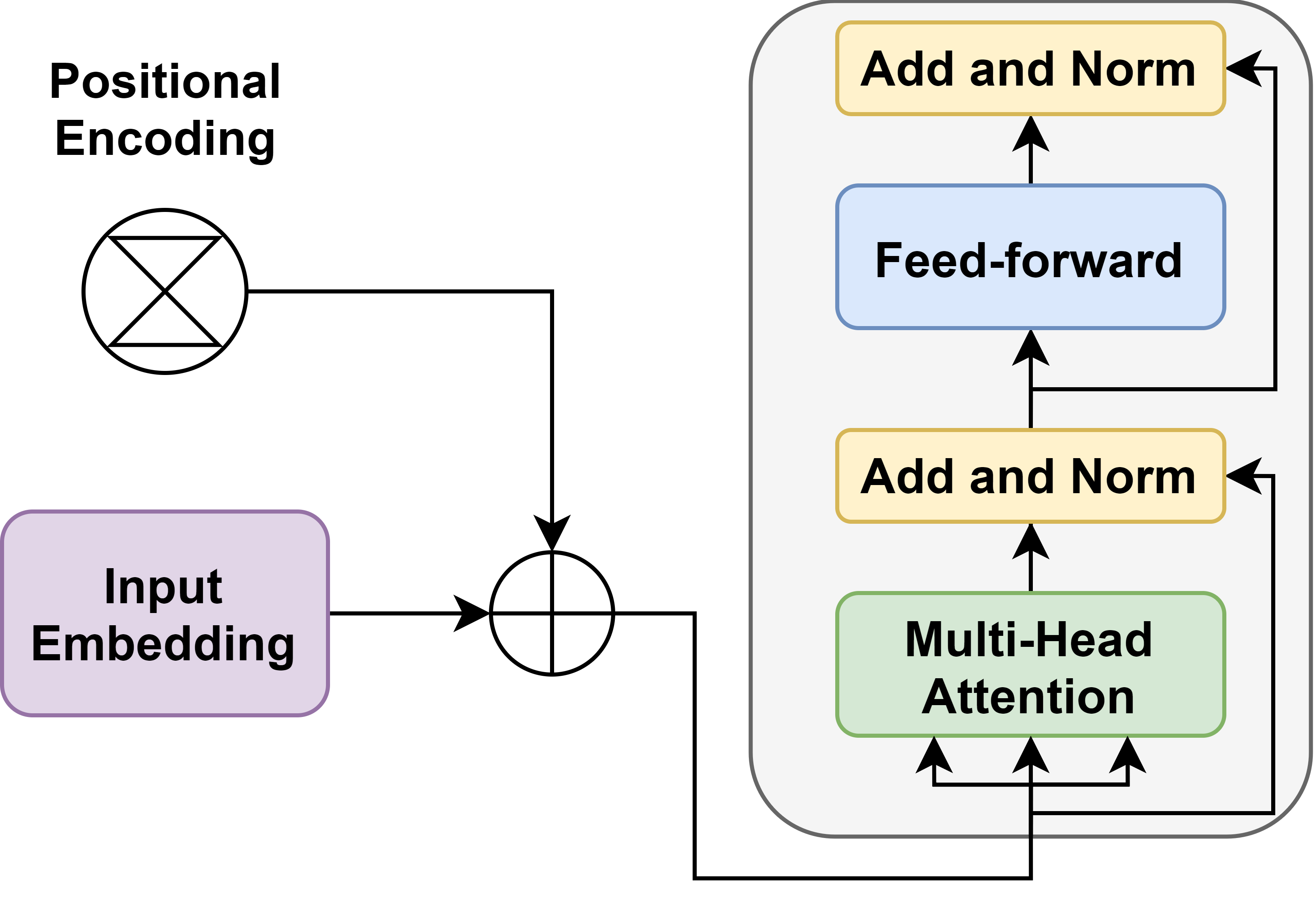}}
\caption{Transformers' encoder. \textit{Add} stands for adding a residual connection \cite{He2015} and \textit{Norm} stands for normalization \cite{ba2016layer}. As can be seen, inside the Transformer encoder, there is also a position-wise fully connected feed-forward network.}
\label{fig:transformer_encoder}
\end{figure}

\subsection{Transformer}

Transformers are architectures developed to process sequential signals and represent the state-of-the-art in NLP tasks. Even though recent studies have increased the range of applications for this architecture, ranging from computer vision \cite{Dosovitskiy2020} to drug response prediction \cite{Ott2018}, Transformer models are still not used with temporal signals of multiple channels, such as EEG. When EEG processing is compared with NLP tasks, in both cases, the data are sequential, and events that occur in a short-term segment need to be contextualized with the rest of the data for its complete understanding \cite{Bertasius2021}.

These architectures are based only on attention mechanisms, dispensing with any convolution or recurrence. Through an encoder and a decoder, this architecture will employ an attention mechanism to focus on the regions of most significant interest for a given input and, consequently, spend a more extensive computational resource in this area. The attention mechanism used by Vaswani et al. \cite{Vaswani2017} was the Scaled Dot-Product Attention, given by: 

\begin{equation}\label{eq:attention}
    Attention(Q,K,V) = softmax(\dfrac{QK^{T}}{\sqrt{d_{k}}})V,
\end{equation}
where $Q$, $K$, and $V$ are vectors called query, key, and value, respectively, that are going to be used inside attention layers and $\sqrt{d_{k}}$ is the so-called scale factor.

In the Transformer encoder (Fig. \ref{fig:transformer_encoder}), the input is provided to the architecture after going through an embedding to convert each input element to vectors of the same dimension. After that, since this model does not use convolution or recurrence, position information for each component will be added to the input via a positional encoding. Vaswani et al. \cite{Vaswani2017} employed attention in different positions of different representations of input subspaces within the encoder through a mechanism called Multi-Head Attention, which allows parallel computation and thus requires less training time.

The advantages of using Transformers are the ability to perform parallel computing and fast training time, at the cost of not supporting large input sequences since the attention mechanisms scale quadratically with the input length \cite{Vaswani2017}.

\subsection{Vision Transformer}

ViT is a Transformer model adapted to use images as input. Thus, instead of processing 1D sequential data, ViT will use 2D images as input. In a first step, ViT will subdivide the input into patches. Then, a linear embedding sequence of these patches and position embeddings are provided as input to a Transformer encoder (see Fig. \ref{fig:transformer_encoder}). While the position embedding adds input topology information, the ViT processes the image with a linear projection of the flattened patches, whose components indicate low-dimensional correlations in the patches, and the Multi-Head Attention mechanism aggregates image information across all layers.

\section{Models} \label{sec:models}

The following section describes DL models and architectures developed here to predict an epileptic seizure using EEG signals as input. In all models, the last layer is composed of one neuron with a sigmoid activation function to perform the binary classification of the EEG periods. Also, each model employed a binary cross-entropy loss function (\ref{eq:binary_loss}). 

\begin{equation}\label{eq:binary_loss}
    l(y,\hat{y}) = - [y \cdot \mathrm{log}(\hat{y}) + (1 - y) \cdot \mathrm{log}(1-\hat{y})]
\end{equation}
Where $l$ is the loss function, $y$ is the desired output, and $\hat{y}$ is the obtained output.
All models used Adam as the optimizer \cite{Kingma2015} with a learning rate of 0.001 and an exponential decay at a rate of 0.94 for each step of $n/128$, where $n$ is the number of samples.

\subsection{Multi-Layer Perceptron}

\begin{figure}[t!]
\centerline{\includegraphics[width=0.6\columnwidth]{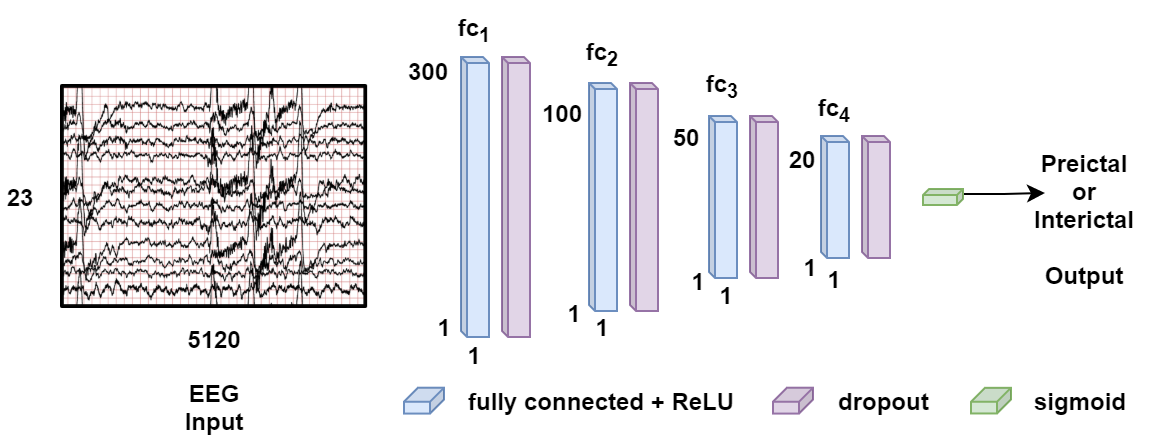}}
\caption{MLP Model. Here, the input is EEG signals with 23 channels and 5,120 time steps, corresponding to 20-second samples at a sampling rate of 256 Hz. In the case of 5-second samples, the input would be 23 channels and 1,280 time steps.}
\label{fig:mlp}
\end{figure}

The first technique implemented is the MLP. It is one of the most common applications of artificial neural networks (ANN), and it is composed of three layers types: input layer, hidden layers, and output layer. These architectures encompass units called neurons that are connected. Thus, when providing data to an MLP, neurons will incorporate information and transmit it to others.

The proposed MLP model consisted of four hidden layers with 300, 100, 50, and 20 neurons, in that order. In each intermediate layer, a ReLU activation function is used to add non-linearity and ensure noise robustness in the input data \cite{Daoud2019}. After each hidden layer, the dropout regularization technique \cite{Srivastava2014} with a value of 0.3 reduces overfitting. This model is illustrated in Fig. \ref{fig:mlp}.

\subsection{CNN with 1D and 2D kernels}

\begin{figure*}[!t]
\centerline{\includegraphics[width=0.90\textwidth]{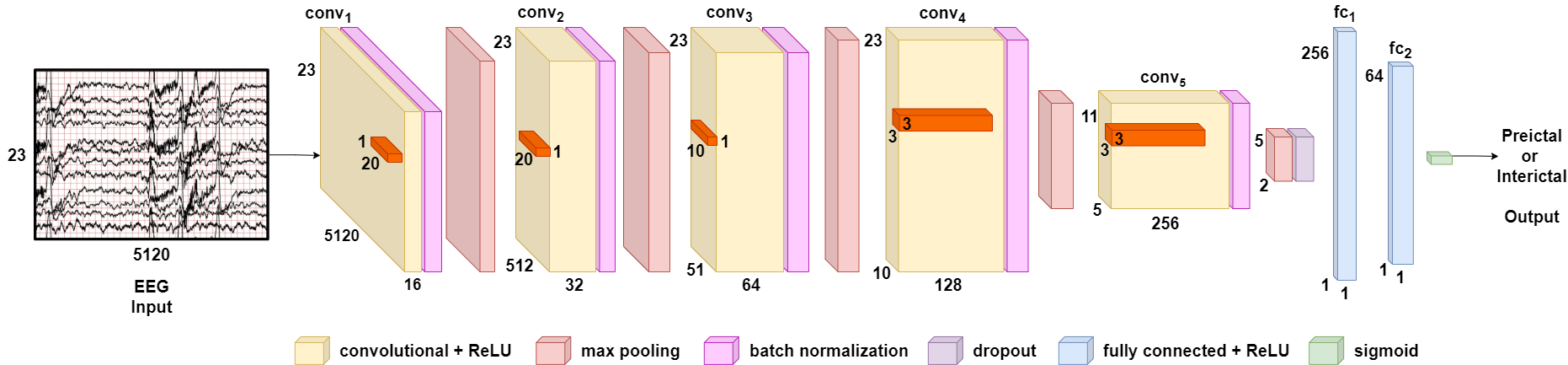}}
\caption{CNN Model. Again, the input is EEG signals with 23 channels and 5,120 time steps. Input dimensions will be gradually reduced through max-pooling layers, while the relevant input information is maintained. Filters are shown in the figure in orange.}
\label{fig:cnn}
\end{figure*}

An end-to-end architecture based on a CNN with 1D and 2D kernels was designed for the seizure prediction task. The justification for using a 1D filter is that, unlike images, EEG signals present little redundancy between the vertical axes, i.e., the electrode channels, since the horizontal axes contain millions of elements and the vertical axis only a few channels. Thus, early 1D filters will learn time-domain characteristics while a max-pooling layer will keep important information. The following 2D filters will learn features in both time and space domains. 

The CNN model developed here is composed of five convolution layers, where the convolution layers are followed by a batch normalization layer \cite{Ioffe2015}, a ReLU activation function, a max-pooling layer, and fully connected layers. The first two convolution blocks employ $1\times20$ size filters and a $1\times10$ max-pooling. The third convolution block employs $1\times10$ size filters and $1\times5$ max-pooling. The last two blocks employ $3\times3$ filters and $2\times2$ max-pooling. The number of filters in the five layers are, in this order, 16, 32, 64, 128, and 256. At the end of the last convolution layer, a dropout value of 0.5 was applied. Finally, the two fully connected layers have 256 and 64 neurons with ReLU activation functions. This model is illustrated in Fig. \ref{fig:cnn}.

\subsection{CNN+Bi-LSTM}
\begin{figure*}[!t]
\centerline{\includegraphics[width=0.89\textwidth]{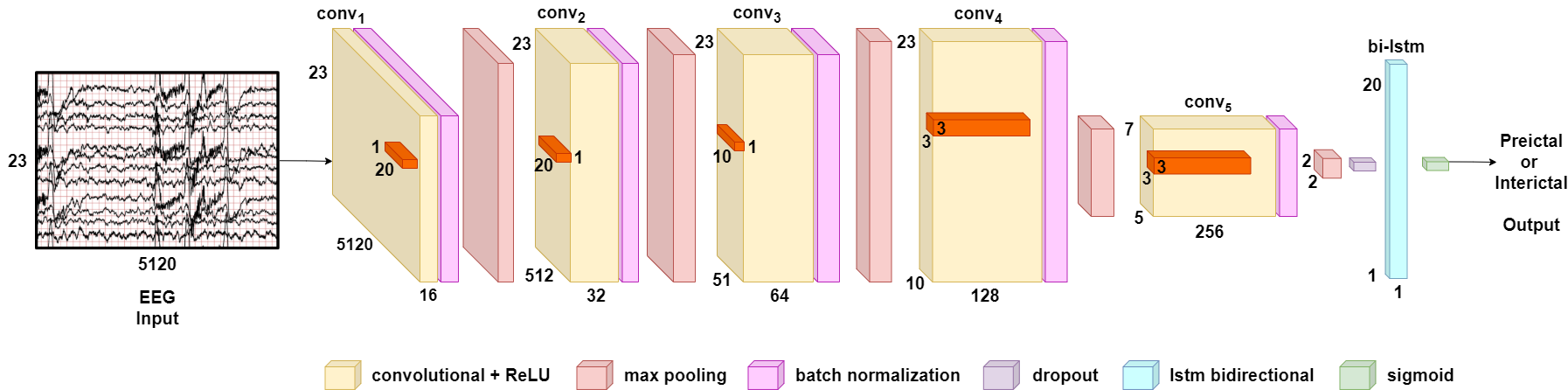}}
\caption{CNN+Bi-LSTM Model. The dimension of the Bi-LSTM layer is 20 since bidirectional networks are nothing more than two networks processing data in opposite directions. As each LSTM has 10 neurons, the Bi-LSTM layer has 20 in total.}
\label{fig:bi-lstm}
\end{figure*}

A DL architecture was developed based on a CNN as the front-end and a Bi-LSTM as the back-end. The CNN will extract the spatial characteristics between the different electrode channels to distinguish the preictal and interictal states. The addition of a Bi-LSTM network aims to extract the temporal features of the input sequence.

This architecture is widely used for temporal data \cite{Yu2019}, and employing a bidirectional network allows to incorporate information from the future states to obtain the current state output during training.

An architecture composed by 5 convolution layers was designed and implemented with number of filters 16, 32, 64, 128, and 256 and filters of dimensions $1\times20$, $1\times20$, $1\times10 $, $3\times3$, and $3\times3$. Each convolution layer is followed by a batch normalization layer and a max-pooling layer of dimensions $1\times10$, $1\times10$, $1\times5$, $3\times2$, and $3\times2$. After the last layer of max-pooling, a flatten layer is employed. Finally, a Bi-LSTM layer of 10 units is added, employing dropout on the input of 0.1 and recurrent dropout of 0.5. This model is illustrated in Fig. \ref{fig:bi-lstm}.

\subsection{Temporal Multi-Channel Transformer: A Transformer adaptation}

\begin{figure*}[!t]
\centerline{\includegraphics[width=0.83\textwidth]{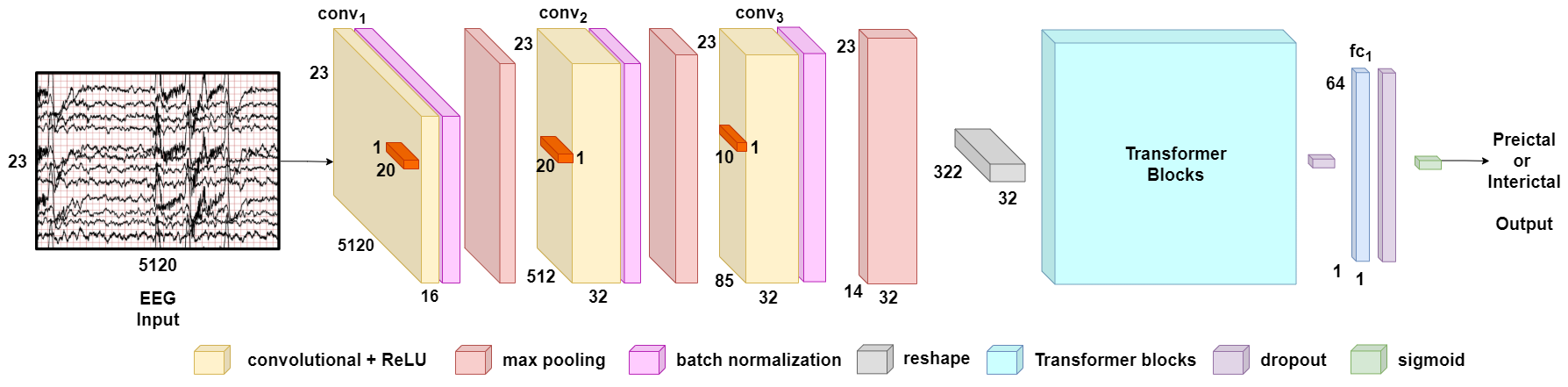}}
\caption{TMC-T Model. Three convolution blocks are used to reduce the input dimensions and extract the embeddings intelligently. After that, the result of the convolutions is flattened and supplied to Transformers' blocks.}
\label{fig:transformer}
\end{figure*}

We presented the model named TMC-T, a variation of the Transformer architecture adapted to process temporal signals with one or more channels. A CNN network was used on the input data before being supplied to the Transformer to make this possible. First, a Transformer architecture was designed with eight attention heads and hidden layers of 64 units in feedforward networks within the Transformer.
For position encoding, learnable embeddings were used. For token embedding, a convolutional network was used. Using a CNN as an embedding has two advantages: 1) Reducing the input size, as each EEG sample to be supplied to the network has 23 lines (number of channels) and 5,120 columns (20 seconds samples acquired at 256 Hz). Since the Transformers do not support long strings as input, the use of CNN followed by max-pooling layers allowed to reduce the input dimension from $23\times5120$ to 322; 2) Extraction of features intelligently. In this architecture, an embedding dimension of 32 was used. Thus, the last convolution layer contains 32 filters that will be the embedding dimension.

The CNN is composed of convolution layers followed by batch normalization and max-pooling layers. There are 3 convolution layers of 16, 32, and 32 filters of dimensions $1\times20$, $1\times20$, and $1\times10$. The max-pooling layers have dimensions of $1\times10$, $1\times6$, and $1\times6$.

A residual dropout is applied to the output of each Transformer sub-layer. The last two dense layers, which carry out the classification, have a dropout value equal to 0.5. Also, a dropout of 0.1 is added to the sum of embeddings and positional encoding in both encoder and decoder. This model is illustrated in Fig. \ref{fig:transformer}.

\subsection{Temporal Multi-Channel Vision Transformer: A Vision Transformer adaptation}

\begin{figure*}[!t]
\centerline{\includegraphics[width=\textwidth]{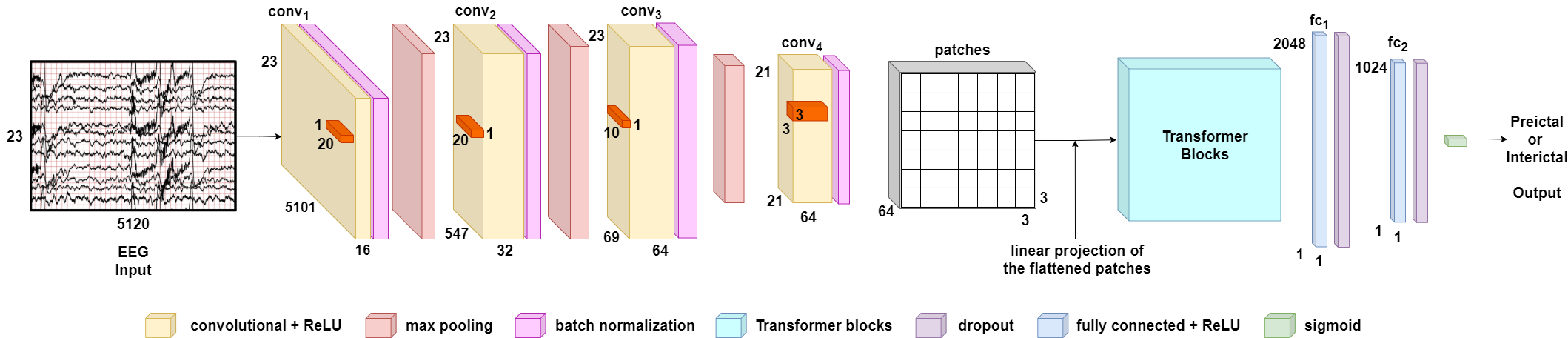}}
\caption{TMC-ViT Model. Four convolution layers will extract the embeddings and reduce the input dimension to a 2D data grid of $21 \times 21$, which will be interpreted as images by ViT and split into patches of $3 \times 3$.}
\label{fig:vit}
\end{figure*}

Dosovitskiy et al. \cite{Dosovitskiy2020} employed this architecture for classifying $16\times16$ images. Thus, to use this network to classify multi-channel EEG signals, a CNN was used again at the input to create the embeddings and reduce the input matrix to dimensions $21\times21$. The choice of this dimension was so that a value close to the images used in the original work, i.e., $16\times16$, was obtained, and patches of size $3\times3$ could be used. Thus, sequential signals from multiple channels will be interpreted as 2D images. This model was named TMC-ViT.

The CNN mentioned above has 16, 32, 64, and 64 filters of dimensions $1\times20$, $1\times20$, $1\times10$, and $3\times3$. Each convolution layer is followed by a batch normalization and max-pooling layers of dimensions $1\times9$, $1\times7$, and $1\times3$.

Again, learnable embeddings were used for the position encoding, and a convolutional network was used for the tokens embedding. Since the last layer of the convolution has 64 filters, embedding has a dimension of 64. The number of attention heads and Transformer layers adopted was 4 and 8, respectively. The last dense layers have dimensions 2,048 and 1,024. This model is illustrated in Fig. \ref{fig:vit}.

\section{Experiments} \label{sec:experiments}

\subsection{CHB-MIT Scalp EEG Database}

The CHB-MIT Scalp EEG Database \cite{Goldberger2000} is a publicly available EEG database obtained from Children's Hospital Boston patients. Seizures (i.e., ictal state) are already adequately identified in this database. These EEG signals were measured in pediatric patients with intractable seizures. To evaluate the possibility of neurosurgery, these patients were monitored for days without antiseizure medications so that their seizures could be observed and further characterized.d.

In total, there are 23 cases in 22 patients, with the subjects' distribution shown in Table \ref{tab:patients}. The CHB-MIT database consists of signals from patients ranging in age from 1.5 to 22 years. Therefore, this dataset contains EEG signals from both pediatric and adult patients. Signals were measured through electrodes placed on the scalp of patients using the 10-20 system at a rate of 256 samples per second, i.e., 256 Hz, with 16-bit resolution. To assess the performance of the five developed models in this paper, we only considered patients with a fixed number of electrodes at 23 and at least three lead seizures, i.e., three or more seizures that occur at least four hours after previous seizures. Moreover, subject chb06 was excluded due to a lack of comparison with recent works \cite{Zhao2020}. The TMC-ViT model, which achieved the best performance among the tested models, was further evaluated for 22 of 23 cases from the CHB-MIT database. The subject chb12 was excluded due to a lack of interictal period samples.

\begin{table}[h]
\centering
\caption{Information on selected patients from the CHB-MIT database.}
\label{tab:patients}
\ra{1.3}
\resizebox{0.7\linewidth}{!}{
\setlength\tabcolsep{2pt}
\begin{tabular}{|c||ccccccccccccccccccccccc|}
\hline
\textbf{\begin{tabular}[c]{@{}c@{}}Case\\ (chb)\end{tabular}} & 01 & 02 & 03 & 04 & 05 & 06 & 07 & 08 & 09 & 10 & 11 & 12 & 13 & 14 & 15 & 16 & 17 & 18 & 19 & 20 & 21 & 22 & 23 \\ \hline
\textbf{Gender} & F & M & F & M & F & F & F & M & F & M & F & F & F & F & M & F & F & F & F & F & F & F & F \\ \hline
\textbf{\begin{tabular}[c]{@{}c@{}}Age\\ (year)\end{tabular}} & 11 & 11 & 14 & 22 & 7 & 1.5 & 14.5 & 3.5 & 10 & 3 & 12 & 2 & 3 & 9 & 16 & 7 & 12 & 18 & 19 & 6 & 13 & 8 & 6 \\ \hline
\textbf{\begin{tabular}[c]{@{}c@{}}No. of\\ Seizures\end{tabular}} & 7 & 3 & 7 & 4 & 5 & 10 & 3 & 5 & 4 & 7 & 3 & 40 & 12 & 8 & 20 & 10 & 3 & 6 & 3 & 8 & 4 & 3 & 7 \\ \hline
\end{tabular}}
\end{table}

\subsection{Preprocessing}\label{subsec:preprocess}
\subsubsection{Preictal and interictal state}
To address questions that are still open in the literature, such as the definition of the preictal period, the definitions of 30 and 60 minutes before a seizure were used to assess which obtains the best result. The interictal state was defined as the time period between 4 hours after the last seizure and 4 hours before the subsequent seizure. Moreover, we defined an SPH of 5 minutes before a seizure. This period is intended to 1) ensure that the medical staff receives the alert of an approaching seizure early enough to take the necessary steps to prepare the patient, and 2) secure that any ictal activity occurring before seizure demarcation that the neurophysiologist may have misclassified does not interfere with the preictal state samples \cite{Benjamin2016}.

\subsubsection{Window size}
Two sizes of EEG samples to be provided for the DL architectures were tested: 5-second samples and 20-second samples, with the 20-second samples using a 5-second overlap, i.e., a segment of the preictal period used for the network training will contain 5 seconds from the end of the previous period and the beginning of the next segment. 

\subsubsection{Data balance}
Since the number of interictal period segments is usually more extensive than the preictal period, the use of overlap only for preictal samples decreases the gap between the number of occurrences in the two periods. In addition, the same number of interictal and preictal samples were provided to our DL models by downsampling the number of interictal cases for the entire training set.

\subsubsection{Raw data}
We employ raw data relying on DL algorithms' ability to automatically learn discriminative features even when data possibly contaminated with noise and artifacts are used as input. By not employing hand-extracted features, we provide data with all information available to the models to identify relevant patterns that the use of feature engineering could miss. The automatic feature extraction of EEG signals via DL proves to be more robust and with more potential than those hand-crafted features \cite{Antoniades2016}. Moreover, we applied batch normalization after each convolution layer and before the activation function to normalize the data at each mini-batch. The use of batch normalization incorporates normalization within the network architecture itself, eliminating the need to do so in preprocessing \cite{Ioffe2015}.

\subsection{Training and Evaluation}
The models were trained on an AWS EC2 \textit{g4dn.2xlarge} instance with 32GiB of memory, 8vCPUs, and a GPU. One model was trained for each patient, so the solutions developed here are patient-specific. Since patients with epilepsy are usually monitored for days without antiseizure medications so that their seizures can be observed and characterized, patient-specific solutions prove to be a viable option since a large amount of EEG data from each patient is available for training the models. Also, patient-specific solutions generally perform better than patient-generic. This better performance is since EEG signals present unique characteristics to the patient from which they were collected. The train set corresponds to 80\% of the EEG data of the respective patient after balancing, and the test set corresponds to 20\% of the remaining data. Among the train set, we executed a 5-fold cross-validation with 10\% of the available training data for hyperparameter optimization by empirical evaluation.

To assess the functionality of the algorithm, these widely used metrics in epileptic seizure prediction \cite{Rasheed2021} were evaluated: accuracy (Acc), sensitivity (SS) or recall, specificity (SP), respectively (\ref{eq:acc}), (\ref{eq:ss}), and (\ref{eq:sp}), and area under the curve (AUC), which evaluates sensitivity as a function of the false positive rate at different test thresholds. 

\begin{equation}
    Acc = \frac{TP+TN}{TP+TN+FP+FN}
    \label{eq:acc}
\end{equation}

\begin{equation}
    SS = \frac{TP}{TP+FN}
    \label{eq:ss}
\end{equation}

\begin{equation}
    SP = \frac{TN}{TN+FP}
    \label{eq:sp}
\end{equation}

TP, TN, FP, and FN are true positive, true negative, false positive, and false negative, respectively.

Accuracy was assessed as we balanced the database before models training. Thus, we ensure that our models are not misclassifying all samples as preictal, which would generate high sensitivity rates and low false prediction rate (FPR). The joint analysis of accuracy, sensitivity, specificity, and AUC certifies that we avoid false positives and negatives. 

The majority of authors in the literature do not explain how they calculate the FPR per hour. There is evidence that this metric has been wrongly calculated in the past \cite{Mormann2007}, i.e., uncorrected FPR including preictal periods (every alarm is a true prediction). Therefore, we did not use this metric to avoid the results obtained in this paper compared with other works that miscalculate it.

\section{Results} \label{sec:results}

Each model presented in Section \ref{sec:models} was trained and tested for all six subjects in a patient-specific way using preictal period definitions and EEG sample sizes. The experiments presented in this work employed minimal preprocessing. The use of batch normalization was the critical factor that enabled the use of raw data without normalization steps during preprocessing and reduced training time. We present the results in two subsections: Section \ref{subsec:model_optim}shows the values obtained after hyperparameter optimization for comparison between models when varying preictal and sample sizes, whereas Section \ref{subsec:comparison_literature} provides a comparison with the literature.


\subsection{Comparison between DL techniques}\label{subsec:model_optim}

\begin{table*}
\centering
\caption{The table contains the average of the results obtained between all the six patients by the models of MLP, CNN, CNN+Bi-LSTM, TMC-T, and TMC-ViT for all preictal and sample sizes.}
\label{tab:all_results}
\resizebox{0.9\linewidth}{!}{
\begin{tabular}{ccccccc} 
\toprule
\textbf{Model} & \begin{tabular}[c]{@{}c@{}}\textbf{Preictal}\\\textbf{size (min)}\end{tabular} & \begin{tabular}[c]{@{}c@{}}\textbf{Sample}\\\textbf{size (ms)}\end{tabular} & \textbf{Accuracy (\%)} & \textbf{AUC (\%)} & \textbf{Sensibility (\%)} & \textbf{Specificity (\%)} \\ 
\midrule\midrule
\multirow{4}{*}{\textbf{TMC-ViT}} & \multirow{2}{*}{30} & 5 & $92.49 \pm 7.58$ & $94.99 \pm 5.54$ & $93.11 \pm 7.47$ & $93.17 \pm 7.30$ \\ 
\cmidrule{3-7}
 &  & 20 & $90.69 \pm 6.03$ & $94.96 \pm 4.42$ & $89.71 \pm 7.54$ & $92.17 \pm 7.84$ \\ 
\cmidrule{2-7}
 & \multirow{2}{*}{\textbf{60}} & 5 & $92.31 \pm 7.86$ & $94.39 \pm 5.95$ & $95.52 \pm 4.77$ & $90.97 \pm 10.84$ \\ 
\cmidrule{3-7}
 &  & \textbf{20} & $\mathbf{95.73 \pm 3.56}$ & $\mathbf{97.55 \pm 2.50}$ & $\mathbf{96.46 \pm 3.20}$ & $\mathbf{96.09 \pm 4.20}$ \\ 
\midrule
\multirow{4}{*}{\textbf{TMC-T}} & \multirow{2}{*}{30} & 5 & $88.91 \pm 10.17$ & $91.67 \pm 8.85$ & $90.27 \pm 10.30$ & $89.78 \pm 10.33$ \\ 
\cmidrule{3-7}
 &  & 20 & $89.44 \pm 7.73$ & $92.83 \pm 6.85$ & $87.97 \pm 8.31$ & $91.71 \pm 6.37$ \\ 
\cmidrule{2-7}
 & \multirow{2}{*}{\textbf{60}} & 5 & $90.71 \pm 8.20$ & $93.14 \pm 6.65$ & $92.16 \pm 8.25$ & $91.10 \pm 9.06$ \\
 \cmidrule{3-7}
 &  & \textbf{20} & $\mathbf{93.74 \pm 5.45}$ & $\mathbf{96.26 \pm 4.41}$ & $\mathbf{93.87 \pm 3.68}$ & $\mathbf{93.72 \pm 7.63}$ \\ 
\midrule
\multirow{4}{*}{\textbf{CNN}} & \multirow{2}{*}{30} & 5 & $93.47 \pm 4.93$ & $95.77 \pm 3.58$ & $95.16 \pm 3.89$ & $93.55 \pm 4.84$ \\ 
\cmidrule{3-7}
 &  & 20 & $87.86 \pm 8.72$ & $92.31 \pm 7.84$ & $89.82 \pm 9.02$ & $86.15 \pm 15.00$ \\ 
\cmidrule{2-7}
 & \multirow{2}{*}{\textbf{60}} & \textbf{5} & $\mathbf{95.59 \pm 4.36}$ & $\mathbf{97.31 \pm 2.92}$ & $\mathbf{95.80 \pm 5.17}$ & $\mathbf{96.46 \pm 2.96}$ \\ 
\cmidrule{3-7}
 &  & 20 & $93.86 \pm 3.83$ & $96.83 \pm 2.39$ & $93.95 \pm 4.58$ & $94.72 \pm 3.71$ \\ 
\midrule
\multirow{4}{*}{\begin{tabular}[c]{@{}c@{}}\textbf{CNN}\\+\\\textbf{Bi-LSTM}\end{tabular}} & \multirow{2}{*}{30} & 5 & $88.57 \pm 8.77$ & $92.63 \pm 6.22$ & $90.62 \pm 7.47$ & $89.29 \pm 8.69$ \\ 
\cmidrule{3-7}
 &  & 20 & $89.20 \pm 10.41$ & $93.68 \pm 7.31$ & $92.20 \pm 7.54$ & $91.17 \pm 8.39$ \\ 
\cmidrule{2-7}
 & \multirow{2}{*}{\textbf{60}} & 5 & $91.11 \pm 8.00$ & $94.25 \pm 5.61$ & $92.43 \pm 6.97$ & $90.99 \pm 8.83$ \\ 
\cmidrule{3-7}
 &  & \textbf{20} & $\mathbf{92.07 \pm 7.09}$ & $\mathbf{95.51 \pm 4.84}$ & $\mathbf{94.25 \pm 4.63}$ & $\mathbf{92.11 \pm 7.94}$ \\ 
\midrule
\multirow{4}{*}{\textbf{MLP}} & \multirow{2}{*}{30} & 5 & $74.03 \pm 11.91$ & $78.17 \pm 12.52$ & $74.10 \pm 11.50$ & $74.28 \pm 11.87$ \\ 
\cmidrule{3-7}
 &  & 20 & $64.63 \pm 8.12$ & $67.84 \pm 8.75$ & $61.95 \pm 16.09$ & $61.79 \pm 11.65$ \\ 
\cmidrule{2-7}
 & \multirow{2}{*}{\textbf{60}} & \textbf{5} & $\mathbf{75.02 \pm 11.87}$ & $\mathbf{78.85 \pm 11.24}$ & $\mathbf{77.82 \pm 11.96}$ & $\mathbf{74.20 \pm 9.82}$ \\ 
\cmidrule{3-7}
 &  & 20 & $65.63 \pm 7.86$ & $68.93 \pm 8.07$ & $64.09 \pm 16.80$ & $61.94 \pm 10.68$ \\
\bottomrule
\end{tabular}}
\end{table*}

\begin{figure}
\centerline{\includegraphics[width=0.70\columnwidth]{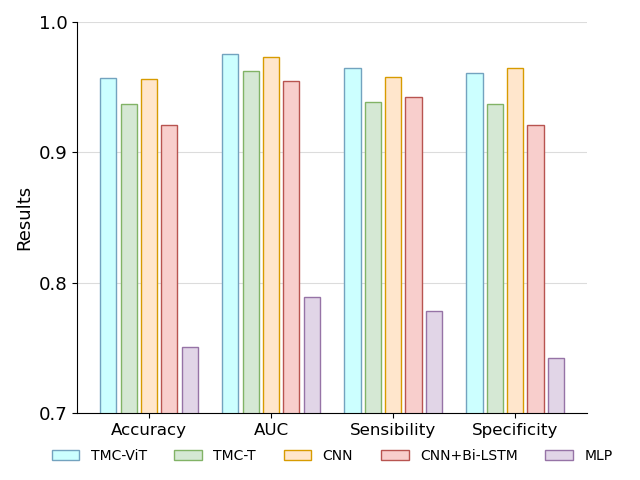}}
\caption{Comparison between the evaluated metrics obtained by each of the implemented models. Values are an average of the results obtained by the models for each of the patients.}
\label{fig:bestmodels}
\end{figure}

\begin{figure}
\centerline{\includegraphics[width=0.70\columnwidth]{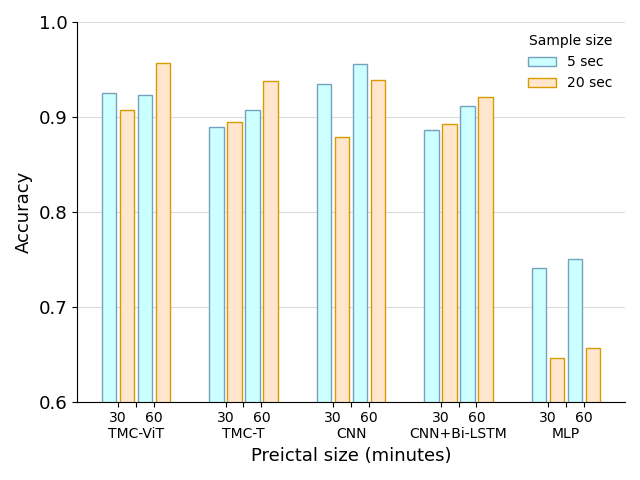}}
\caption{Accuracy of each of the models for both preictal and sample size definitions. It can be noticed that even the worst TMC-ViT model, i.e., 30 minute preictal and 20-second samples, obtain competitive or even superior results than the best models of the other architectures.}
\label{fig:preictal_size}
\end{figure}


The Table \ref{tab:all_results} presents the results achieved by the MLP, CNN, CNN+Bi-LSTM, TMC-T, and TMC-ViT models for both definitions of preictal state and sample sizes. The metrics values are the mean and standard deviation between the 5-folds for all tested patients. 
The best model of each DL technique, except MLP, achieved results above 90\% for all evaluated metrics, as shown in Fig. \ref{fig:bestmodels}. Our TMC-ViT obtained the highest results for accuracy (95.73\%), AUC (97.55\%), and sensibility (96.46\%). Furthermore, it achieved the lowest standard deviation among patients for the mentioned metrics, indicating the model's robustness in learning the nuances of EEG signals from different types of seizures and patients.

The CNN architecture achieved the best specificity with 96.46\% and the lowest standard deviation for this metric, whereas the TMC-ViT model achieved 96.09\%. The MLP obtained an accuracy of around 75\% and AUC above 78\%, which was expected since this architecture is the simplest compared to the others, thus with less potential for adaptation to the nuances of each patient using only raw EEG signals as input.


Concerning the EEG samples provided to the networks, the use of 20-second segments with a 5-second overlap for the preictal period proved to be beneficial for the TMC-ViT, TMC-T, and CNN+Bi-LSTM models, as shown in Fig. \ref{fig:preictal_size}. 
We can observe a gradual gain in accuracy in these models, as in other evaluated metrics (see Table \ref{tab:all_results}), when we increase the preictal period and sample sizes. This implies two conclusions: 

\begin{enumerate}
    \item The use of overlap for preictal samples reduced the gap in the number of samples between the preictal and interictal period improving the results, indicating that the fewer number of preictal samples is indeed one of the biggest challenges in the task of predicting epileptic seizures;
    \item These models could benefit from larger samples, being able to incorporate long-term features within the sample efficiently.
\end{enumerate}


When employing a Bi-LSTM layer at the end of the CNN, the only case in the results was better than the CNN model with fully connected layers was for samples of 20 seconds and a preictal size of 30 minutes. In this case, the Bi-LSTM benefited from larger samples due to its ability to learn long-term dependencies within the sample. However, when testing for a preictal of 60 min, with a larger number of samples, the model was significantly slower for training and did not improve the overall results at the same pace as CNN or Transformer-based networks.

Fig. \ref{fig:preictal_size} also highlights the inefficiency of the MLP model in dealing with large input samples, with 20-second samples being worse than 5-second samples for both preictal definitions. The overlap in the MLP model resulted in high overfitting during training. The MLP accuracy for a preictal of 60 minutes, for example, is 75\% for 5-second samples against 65.63\% for 20-second samples, i.e., a reduction of almost 10\%. 

A model employing shorter samples is a positive aspect, especially in online applications where sample classification time is paramount. In contrast, the fact that CNN has shown better results for 5-second samples without overlap is not due to an inefficiency in dealing with larger samples but to the efficiency in learning the relevant features of smaller samples without the need for overlap, and hence fewer training samples. CNN's ability to learn optimal parameters even from shorter samples could result from its weight sharing and its smaller number of learnable parameters.

All models showed better results for a preictal definition of 60 minutes. It indicates that the DL models tested can identify relevant patterns of the preictal samples far in time, even though the 60-minute definition may cover noisier samples as it moves further away from the period of occurrence of the seizure, i.e., the ictal state. For this definition, the best TMC-ViT model employed samples of 20 seconds with overlap, whereas for CNN, the best model used samples of 5 seconds. The TMC-ViT model surpassed CNN in three of the four metrics evaluated, by employing overlap in the samples and consequently increasing the dataset. The tendency of ViT models to outperform CNN in more extensive databases was also observed in \cite{Dosovitskiy2020}.

Finally, we used McNamer's test \cite{Kim2017} to compare these two candidates: TMC-ViT with a 20-second window and CNN with a 5-second window. Since they have different windows sizes, we run the prediction with CNN and average the output for four consecutive windows, thus, generating an equivalent 20-second prediction. This way, we were able to compare both models by rejecting the null hypothesis of no difference in performance between the models, with a p-value ($p=0.0064$) smaller than the significance threshold ($\alpha < 0.05$). Therefore, the TMC-ViT model performs better for the evaluated samples.





\subsection{Comparison with literature}\label{subsec:comparison_literature}

\begin{table*}[!h]
\centering
\caption{Comparison of the results obtained for each patient by the TMC-ViT model with the literature.}
\label{tab:results_subjects}
\resizebox{\textwidth}{!}{
\setlength\tabcolsep{3pt}
\begin{tabular}{cccccccccccc} 
\toprule
\multirow{2}{*}{\textbf{Subject}} & \multicolumn{3}{c}{\textbf{Acuraccy (\%)}} & \multicolumn{3}{c}{\textbf{AUC (\%)}} & \multicolumn{4}{c}{\textbf{Sensibility (\%)}} & \textbf{Specificity (\%)} \\ 
\cmidrule{2-12}
 & \begin{tabular}[c]{@{}c@{}}\textbf{TMC-ViT}\\\textbf{(this work)}\end{tabular} & \begin{tabular}[c]{@{}c@{}}CNN+ELM\\\cite{Qin2020}\end{tabular} & \begin{tabular}[c]{@{}c@{}}CNN\\\cite{Zhang2020}\end{tabular} & \begin{tabular}[c]{@{}c@{}}\textbf{TMC-ViT}\\\textbf{(this work)}\end{tabular} & \begin{tabular}[c]{@{}c@{}}BSDCNN\\\cite{Zhao2020}\end{tabular} & \begin{tabular}[c]{@{}c@{}}CNN\\\cite{Zhang2020}\end{tabular} & \begin{tabular}[c]{@{}c@{}}\textbf{TMC-ViT}\\\textbf{(this work)}\end{tabular} & \begin{tabular}[c]{@{}c@{}}CNN+ELM\\\cite{Qin2020}\end{tabular} & \begin{tabular}[c]{@{}c@{}}BSDCNN\\\cite{Zhao2020}\end{tabular} & \begin{tabular}[c]{@{}c@{}}CNN\\\cite{Zhang2020}\end{tabular} & \begin{tabular}[c]{@{}c@{}}\textbf{TMC-ViT}\\\textbf{(this work)}\end{tabular} \\ 
\midrule\midrule
chb01 & \textbf{99.97} & 100 & 94.00 & \textbf{99.97} & 100 & 94.00 & \textbf{100} & 100 & 100 & 92.00 & \textbf{99.94} \\ 
\midrule
chb05 & \textbf{97.16} & 88.88 & 89.00 & \textbf{99.38} & 97.00 & 97.00 & \textbf{96.56} & 89.08 & 90.60 & 90.00 & \textbf{97.94} \\ 
\midrule
chb08 & \textbf{99.10} & 96.96 & 80.00 & \textbf{99.78} & 99.00 & 80.00 & \textbf{99.30} & 97.96 & 99.24 & 78.00 & \textbf{99.11} \\ 
\midrule
chb10 & \textbf{93.48} & 100 & 96.00 & \textbf{96.43} & 96.00 & 96.00 & \textbf{94.17} & 100 & 94.26 & 98.00 & \textbf{93.55} \\ 
\midrule
chb14 & \textbf{90.93} & 90.70 & 84.00 & \textbf{94.01} & 94.00 & 85.00 & \textbf{91.49} & 92.15 & 93.90 & 84.00 & \textbf{88.80} \\ 
\midrule
chb22 & \textbf{93.73} & - & 79.00 & \textbf{95.75} & 93.00 & 80.00 & \textbf{97.23} & - & 93.29 & 83.00 & \textbf{97.23} \\ 
\midrule\midrule
Avarage & \textbf{95.73} & 95.31 & 87.00 & \textbf{97.55} & 97.00 & 88.67 & \textbf{96.46} & 95.83 & 94.69 & 87.50 & \textbf{96.09} \\
\bottomrule
\end{tabular}}
\end{table*}


The TMC-ViT model obtained different results among patients, even though we trained but did not optimize one model per patient. Details are shown in Table \ref{tab:results_subjects}. This is a consequence of the different number and types of seizures and the unique characteristics present in each patient's EEG signal. Our TMC-ViT model obtained consistent results among all patients, with values close to or greater than 90\% for all metrics and evaluated patients, reaching values even of 100\%, as for patient chb01. 



These values are then directly compared with those works in the literature that present the results obtained for each patient \cite{Zhang2020, Zhao2020, Qin2020}. 
Table \ref{tab:results_subjects}shows that our model obtained competitive values in all metrics with the literature for each patient, besides being the only one to make a complimentary analysis of sensitivity and specificity. By maximizing both metrics, instead of just minimizing only false positives like the literature, we also reduce false negatives. When employing the solutions developed here in online applications, our goal is that the patient does not even receive false alarms that a seizure is coming, i.e., a false positive, nor fail to be warned when a seizure is indeed coming, i.e., false negative.

\begin{table}
\centering
\caption{Comparison of the methods used between this work and those in the literature.}
\label{tab:methods_comparison}
\resizebox{0.5\columnwidth}{!}{
\setlength\tabcolsep{3pt}
\begin{tabular}{cccccccc} 
\toprule
\textbf{Model} & \begin{tabular}[c]{@{}c@{}}\textbf{Data} \\\textbf{type}\end{tabular} & \begin{tabular}[c]{@{}c@{}}\textbf{Cross}\\\textbf{Validation}\end{tabular} & \begin{tabular}[c]{@{}c@{}}\textbf{Seizures}\\\textbf{employed}\end{tabular} & \begin{tabular}[c]{@{}c@{}}\textbf{Acc}\\\textbf{(\%)}\end{tabular} & \begin{tabular}[c]{@{}c@{}}\textbf{AUC}\\\textbf{(\%)}\end{tabular} & \begin{tabular}[c]{@{}c@{}}\textbf{SS}\\\textbf{(\%)}\end{tabular} & \begin{tabular}[c]{@{}c@{}}\textbf{SP}\\\textbf{(\%)}\end{tabular} \\ 
\midrule\midrule
\begin{tabular}[c]{@{}c@{}}\textbf{TMC-ViT}\\\textbf{(this work)}\end{tabular} & \textbf{Raw} & \textbf{5-fold} &\textbf{ All} & \textbf{95.73} & \textbf{97.55} & \textbf{96.46} & \textbf{96.09} \\ 
\midrule
\begin{tabular}[c]{@{}c@{}}CNN+ELM\\ \cite{Qin2020} \end{tabular} & STFT & - & All & 95.31 & - & 95.83 & - \\ 
\midrule
\begin{tabular}[c]{@{}c@{}}CNN\\ \cite{Zhang2020}\end{tabular} & WL+CSP & leave-one-out & All & 87.00 & 88.67 & 87.50 & - \\ 
\midrule
\begin{tabular}[c]{@{}c@{}}BSDCNN\\ \cite{Zhao2020} \end{tabular} & Raw & - & - & - & 97.00 & 94.69 & - \\ 
\bottomrule
\end{tabular}}
\end{table}

Table \ref{tab:methods_comparison} presents a comparison between this work and those in the literature concerning the methods used in the experiments and the average results.
Our TMC-ViT model achieved the highest average accuracy and specificity compared to the works of literature. Regarding AUC and sensitivity, our work had competitive results while it is the only one to employ, at the same time: 1) raw data, and consequently minimal preprocessing and less need for prior human knowledge of signals; 2) an algorithm for validation, which ensures the robustness of our algorithms to dataset variations; and 3) to use all seizures for model training and testing so that our solutions can predict any seizure of a given patient, regardless of its duration and period of occurrence.

\subsection{Further evaluation of the TMC-ViT model}

\begin{figure}[!t]
\centerline{\includegraphics[width=0.50\columnwidth]{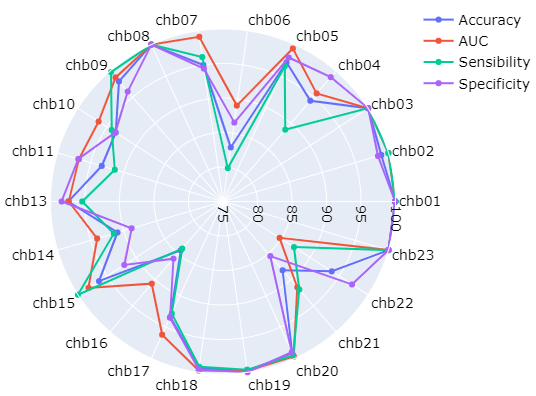}}
\caption{Accuracy, AUC, sensibility, and specificity were achieved by the TMC-ViT model for 22 of 23 subjects from the CHB-MIT dataset.}
\label{fig:all_patients}
\end{figure}

The TMC-ViT model was further evaluated for all the subjects from the CHB-MIT database, except for chb12, due to a lack of interictal period samples. For the models to be trained in all these patients, "dummy", electrocardiogram, and vagal nerve stimulus signals were removed during the preprocessing steps. The results achieved by the TMC-ViT model are shown in Fig. \ref{fig:all_patients}.

The results demonstrate the robustness of our TMC-ViT model. The model achieved over 82\% accuracy, 89\% AUC, 80\% sensitivity, and 86\% specificity for all patients tested. Furthermore, for most patients, the TMC-ViT model reached values equal to or very close to 100\% for all the testes metrics.

\section{Conclusions} \label{sec:conclusion}


This work proposed a novel deep learning architecture to predict epileptic seizures using raw EEG signals. We modified Transformer-based models to deal with temporal series of electrophysiological signals of multiple channels. These models were named TMC-T and TMC-ViT. Additionally, we conducted a thorough analysis regarding sample and preictal sizes, comparing the results with well-established DL techniques. 

Results obtained through computational tests helped clarify questions that are still open related to the definition of the preictal period since its duration does not have a consensus among clinical experts. The two definitions of preictal most used in the literature, 30 and 60 minutes, were tested. It was found that, for raw EEG data and DL models, the 60-minutes preictal definition presented the best results for all the tested models. Concerning sample sizes, the TMC-ViT, TMC-T, and CNN+Bi-LSTM models achieved the highest results when employing 20-second samples, whereas the CNN and MLP models showed better performance for 5-second samples.
The proposed TMC-ViT model surpassed the CNN, which is the current state-of-the-art in predicting epileptic seizures, in three of the four metrics evaluated. Moreover, we compared the TMC-ViT against the CNN model using McNemar's test, showing that the transformer-based algorithm performs better. 
The TMC-ViT model was further evaluated on 22 of 23 subjects from the CHB-MIT database, achieving excellent results among all of them.
The results obtained in this work exceeded several aspects of the literature since the architectures tested here obtained competitive results with the state-of-the-art, employing minimal data preprocessing, raw EEG data as input, a validation algorithm, and predicting all seizures for each patient.

The online prediction of seizures in patients submitted to a long-term video-EEG should be studied for future work. Since epilepsy can be diagnosed by evaluating the patient's cardiovascular, respiratory, and motor changes and cerebral signals \cite{VandeVel2016}, a multimodal framework could be developed using multiple bodily signals from those monitored patients as input to predict epileptic seizures \cite{Rasheed2021}.

\bibliographystyle{unsrtnat}
\bibliography{references}  

\begin{thebibliography}{39}
\providecommand{\natexlab}[1]{#1}
\providecommand{\url}[1]{\texttt{#1}}
\expandafter\ifx\csname urlstyle\endcsname\relax
  \providecommand{\doi}[1]{doi: #1}\else
  \providecommand{\doi}{doi: \begingroup \urlstyle{rm}\Url}\fi

\bibitem[Fisher et~al.(2014)Fisher, Acevedo, Arzimanoglou, Bogacz, Cross,
  Elger, Engel~Jr, Forsgren, French, Glynn, Hesdorffer, Lee, Mathern, Moshé,
  Perucca, Scheffer, Tomson, Watanabe, and Wiebe]{Fisher2014b}
Robert~S. Fisher, Carlos Acevedo, Alexis Arzimanoglou, Alicia Bogacz, J.~Helen
  Cross, Christian~E. Elger, Jerome Engel~Jr, Lars Forsgren, Jacqueline~A.
  French, Mike Glynn, Dale~C. Hesdorffer, B.I. Lee, Gary~W. Mathern, Solomon~L.
  Moshé, Emilio Perucca, Ingrid~E. Scheffer, Torbjörn Tomson, Masako
  Watanabe, and Samuel Wiebe.
\newblock Ilae official report: A practical clinical definition of epilepsy.
\newblock \emph{Epilepsia}, 2014.

\bibitem[Weinstein(2016)]{Weinstein2016}
Steven Weinstein.
\newblock {Seizures and epilepsy: An overview}.
\newblock \emph{Epilepsy: The Intersection of Neurosciences, Biology,
  Mathematics, Engineering, and Physics}, pages 65--77, 2016.

\bibitem[Kwan et~al.(2009)Kwan, Arzimanoglou, Berg, Brodie, {Allen Hauser},
  Mathern, Mosh{\'{e}}, Perucca, Wiebe, and French]{Kwan2009}
Patrick Kwan, Alexis Arzimanoglou, Anne~T Berg, Martin~J Brodie, W~{Allen
  Hauser}, Gary Mathern, Solomon~L Mosh{\'{e}}, Emilio Perucca, Samuel Wiebe,
  and Jacqueline French.
\newblock {Definition of drug resistant epilepsy: Consensus proposal by the ad
  hoc Task Force of the ILAE Commission on Therapeutic Strategies}.
\newblock \emph{Epilepsia}, 51\penalty0 (6):\penalty0 1069--1077, 2009.

\bibitem[459(2020)]{459}
Clinical risk factors in sudep.
\newblock \emph{Neurology}, 94\penalty0 (10), 2020.
\newblock ISSN 0028-3878.

\bibitem[Acharya et~al.(2013)Acharya, {Vinitha Sree}, Swapna, Martis, and
  Suri]{Acharya2013}
U.~Rajendra Acharya, S.~{Vinitha Sree}, G.~Swapna, Roshan~Joy Martis, and
  Jasjit~S. Suri.
\newblock {Automated EEG analysis of epilepsy: A review}.
\newblock \emph{Knowledge-Based Systems}, 45:\penalty0 147--165, 2013.

\bibitem[Chiang et~al.(2011)Chiang, Chang, Chen, Chen, and Chen]{Chiang2011}
Cheng~Yi Chiang, Nai~Fu Chang, Tung~Chien Chen, Hong~Hui Chen, and Liang~Gee
  Chen.
\newblock {Seizure prediction based on classification of EEG synchronization
  patterns with on-line retraining and post-processing scheme}.
\newblock \emph{Proceedings of the Annual International Conference of the IEEE
  Engineering in Medicine and Biology Society, EMBS}, pages 7564--7569, 2011.

\bibitem[Rasheed et~al.(2021)Rasheed, Qayyum, Qadir, Sivathamboo, Kwan,
  Kuhlmann, O'Brien, and Razi]{Rasheed2021}
Khansa Rasheed, Adnan Qayyum, Junaid Qadir, Shobi Sivathamboo, Patrick Kwan,
  Levin Kuhlmann, Terence O'Brien, and Adeel Razi.
\newblock {Machine Learning for Predicting Epileptic Seizures Using EEG
  Signals: A Review}.
\newblock \emph{IEEE Reviews in Biomedical Engineering}, 14:\penalty0 139--155,
  2021.

\bibitem[{Ismail Fawaz} et~al.(2019){Ismail Fawaz}, Forestier, Weber,
  Idoumghar, and Muller]{IsmailFawaz2019}
Hassan {Ismail Fawaz}, Germain Forestier, Jonathan Weber, Lhassane Idoumghar,
  and Pierre~Alain Muller.
\newblock {Deep learning for time series classification: a review}.
\newblock \emph{Data Mining and Knowledge Discovery}, 33\penalty0 (4):\penalty0
  917--963, 2019.

\bibitem[Vaswani et~al.(2017)Vaswani, Shazeer, Parmar, Uszkoreit, Jones, Gomez,
  Kaiser, and Polosukhin]{Vaswani2017}
Ashish Vaswani, Noam Shazeer, Niki Parmar, Jakob Uszkoreit, Llion Jones,
  Aidan~N. Gomez, {\L}ukasz Kaiser, and Illia Polosukhin.
\newblock {Attention is all you need}.
\newblock \emph{Advances in Neural Information Processing Systems}, 2017.

\bibitem[Godoy et~al.(2022{\natexlab{a}})Godoy, Dwivedi, Guan, Turner, Shieff,
  and Liarokapis]{godoy2022access}
Ricardo~V. Godoy, Anany Dwivedi, Bonnie Guan, Amber Turner, Dasha Shieff, and
  Minas Liarokapis.
\newblock On emg based dexterous robotic telemanipulation: Assessing machine
  learning techniques, feature extraction methods, and shared control schemes.
\newblock \emph{IEEE Access}, pages 1--1, 2022{\natexlab{a}}.
\newblock \doi{10.1109/ACCESS.2022.3206436}.

\bibitem[Godoy et~al.(2022{\natexlab{b}})Godoy, Dwivedi, and
  Liarokapis]{godoy2022tnsre}
Ricardo~V. Godoy, Anany Dwivedi, and Minas Liarokapis.
\newblock Electromyography based decoding of dexterous, in-hand manipulation
  motions with temporal multichannel vision transformers.
\newblock \emph{IEEE Transactions on Neural Systems and Rehabilitation
  Engineering}, 30:\penalty0 2207--2216, 2022{\natexlab{b}}.
\newblock \doi{10.1109/TNSRE.2022.3196622}.

\bibitem[Godoy et~al.(2022{\natexlab{c}})Godoy, Lahr, Dwivedi, Reis, Polegato,
  Becker, Caurin, and Liarokapis]{godoy2022ral}
Ricardo~V. Godoy, Gustavo J.~G. Lahr, Anany Dwivedi, Tharik J.~S. Reis,
  Paulo~H. Polegato, Marcelo Becker, Glauco A.~P. Caurin, and Minas Liarokapis.
\newblock Electromyography-based, robust hand motion classification employing
  temporal multi-channel vision transformers.
\newblock \emph{IEEE Robotics and Automation Letters}, 7\penalty0 (4):\penalty0
  10200--10207, 2022{\natexlab{c}}.
\newblock \doi{10.1109/LRA.2022.3192623}.

\bibitem[Goldberger et~al.(2000)Goldberger, Amaral, Glass, Hausdorff, Ivanov,
  Mark, Mietus, Moody, Peng, and Stanley]{Goldberger2000}
A.~L. Goldberger, L.~A. Amaral, L.~Glass, J.~M. Hausdorff, P.~C. Ivanov, R.~G.
  Mark, J.~E. Mietus, G.~B. Moody, C.~K. Peng, and H.~E. Stanley.
\newblock {PhysioBank, PhysioToolkit, and PhysioNet: components of a new
  research resource for complex physiologic signals.}
\newblock \emph{Circulation}, 101\penalty0 (23), 2000.

\bibitem[Daoud and Bayoumi(2019)]{Daoud2019}
Hisham Daoud and Magdy~A. Bayoumi.
\newblock {Efficient Epileptic Seizure Prediction Based on Deep Learning}.
\newblock \emph{IEEE Transactions on Biomedical Circuits and Systems},
  13\penalty0 (5):\penalty0 804--813, 2019.
\newblock ISSN 19409990.

\bibitem[Zhang et~al.(2020)Zhang, Guo, Yang, Chen, and Lo]{Zhang2020}
Yuan Zhang, Yao Guo, Po~Yang, Wei Chen, and Benny Lo.
\newblock {Epilepsy Seizure Prediction on EEG Using Common Spatial Pattern and
  Convolutional Neural Network}.
\newblock \emph{IEEE Journal of Biomedical and Health Informatics}, 24\penalty0
  (2):\penalty0 465--474, 2020.

\bibitem[Zhao et~al.(2020)Zhao, Yang, Xu, and Sawan]{Zhao2020}
Shiqi Zhao, Jie Yang, Yankun Xu, and Mohamad Sawan.
\newblock {Binary Single-Dimensional Convolutional Neural Network for Seizure
  Prediction}.
\newblock pages 1--5, 2020.

\bibitem[Qin et~al.(2020)Qin, Zheng, Chen, Qin, Han, and Che]{Qin2020}
Yingmei Qin, Hailing Zheng, Wei Chen, Qing Qin, Chunxiao Han, and Yanqiu Che.
\newblock {Patient-specific Seizure Prediction with Scalp EEG Using
  Convolutional Neural Network and Extreme Learning Machine}.
\newblock \emph{Chinese Control Conference, CCC}, 2020-July:\penalty0
  7622--7625, 2020.

\bibitem[Yang et~al.(2021)Yang, Zhao, Sun, Lu, and Ma]{Yang2021}
Xinwu Yang, Jiaqi Zhao, Qi~Sun, Jianbo Lu, and Xu~Ma.
\newblock {An effective dual self-attention residual network for seizure
  prediction}.
\newblock \emph{IEEE Transactions on Neural Systems and Rehabilitation
  Engineering}, 2021.

\bibitem[Federico et~al.(2005)Federico, Abbott, Briellmann, Harvey, and
  Jackson]{Federico2005}
Paolo Federico, David~F. Abbott, Regula~S. Briellmann, A.~Simon Harvey, and
  Graeme~D. Jackson.
\newblock {Functional MRI of the pre-ictal state}.
\newblock \emph{Brain}, 128, 2005.
\newblock ISSN 14602156.
\newblock \doi{10.1093/brain/awh533}.

\bibitem[Direito et~al.(2017)Direito, Teixeira, Sales, Castelo-Branco, and
  Dourado]{Direito2017}
Bruno Direito, C{\'{e}}sar~A. Teixeira, Francisco Sales, Miguel Castelo-Branco,
  and Ant{\'{o}}nio Dourado.
\newblock {A Realistic Seizure Prediction Study Based on Multiclass SVM}.
\newblock \emph{International Journal of Neural Systems}, 27\penalty0
  (3):\penalty0 1--15, 2017.
\newblock ISSN 01290657.
\newblock \doi{10.1142/S012906571750006X}.

\bibitem[Teixeira et~al.(2014)Teixeira, Favaro, Direito, Bandarabadi,
  Feldwisch-Drentrup, Ihle, Alvarado, {Le Van Quyen}, Schelter,
  Schulze-Bonhage, Sales, Navarro, and Dourado]{Teixeira2014}
C{\'{e}}sar Teixeira, Gianpietro Favaro, Bruno Direito, Mojtaba Bandarabadi,
  Hinnerk Feldwisch-Drentrup, Matthias Ihle, Catalina Alvarado, Michel {Le Van
  Quyen}, Bjorn Schelter, Andreas Schulze-Bonhage, Francisco Sales, Vincent
  Navarro, and Ant{\'{o}}nio Dourado.
\newblock {Brainatic: A system for real-time epileptic seizure prediction}.
\newblock \emph{Biosystems and Biorobotics}, 6:\penalty0 7--17, 2014.
\newblock ISSN 21953570.
\newblock \doi{10.1007/978-3-642-54707-2_2}.

\bibitem[Ihle et~al.(2012)Ihle, Feldwisch-Drentrup, Teixeira, Witon, Schelter,
  Timmer, and Schulze-Bonhage]{Ihle2012}
Matthias Ihle, Hinnerk Feldwisch-Drentrup, C{\'{e}}sar~A. Teixeira, Adrien
  Witon, Bj{\"{o}}rn Schelter, Jens Timmer, and Andreas Schulze-Bonhage.
\newblock {EPILEPSIAE - A European epilepsy database}.
\newblock \emph{Computer Methods and Programs in Biomedicine}, 106\penalty0
  (3):\penalty0 127--138, 2012.
\newblock ISSN 01692607.
\newblock \doi{10.1016/j.cmpb.2010.08.011}.

\bibitem[Ott et~al.(2018)Ott, Edunov, Grangier, and Auli]{Ott2018}
Myle Ott, Sergey Edunov, David Grangier, and Michael Auli.
\newblock {Scaling neural machine translation}.
\newblock \emph{arXiv}, 2018.

\bibitem[Brown et~al.(2020)Brown, Mann, Ryder, Subbiah, Kaplan, Dhariwal,
  Neelakantan, Shyam, Sastry, Askell, Agarwal, Herbert-Voss, Krueger, Henighan,
  Child, Ramesh, Ziegler, Wu, Winter, Hesse, Chen, Sigler, Litwin, Gray, Chess,
  Clark, Berner, McCandlish, Radford, Sutskever, and Amodei]{Brown2020}
Tom Brown, Benjamin Mann, Nick Ryder, Melanie Subbiah, Jared~D Kaplan, Prafulla
  Dhariwal, Arvind Neelakantan, Pranav Shyam, Girish Sastry, Amanda Askell,
  Sandhini Agarwal, Ariel Herbert-Voss, Gretchen Krueger, Tom Henighan, Rewon
  Child, Aditya Ramesh, Daniel Ziegler, Jeffrey Wu, Clemens Winter, Chris
  Hesse, Mark Chen, Eric Sigler, Mateusz Litwin, Scott Gray, Benjamin Chess,
  Jack Clark, Christopher Berner, Sam McCandlish, Alec Radford, Ilya Sutskever,
  and Dario Amodei.
\newblock Language models are few-shot learners.
\newblock 2020.

\bibitem[Dosovitskiy et~al.(2020)Dosovitskiy, Beyer, Kolesnikov, Weissenborn,
  Zhai, Unterthiner, Dehghani, Minderer, Heigold, Gelly, Uszkoreit, and
  Houlsby]{Dosovitskiy2020}
Alexey Dosovitskiy, Lucas Beyer, Alexander Kolesnikov, Dirk Weissenborn,
  Xiaohua Zhai, Thomas Unterthiner, Mostafa Dehghani, Matthias Minderer, Georg
  Heigold, Sylvain Gelly, Jakob Uszkoreit, and Neil Houlsby.
\newblock {An Image is Worth 16x16 Words: Transformers for Image Recognition at
  Scale}.
\newblock pages 1--21, 2020.

\bibitem[Krishna et~al.(2019)Krishna, Tran, Carnahan, and Tewfik]{Krishna2019}
Gautam Krishna, Co~Tran, Mason Carnahan, and Ahmed~H. Tewfik.
\newblock {EEG based continuous speech recognition using transformers}.
\newblock \emph{arXiv}, 2019.

\bibitem[Choong et~al.(2020)Choong, Hakeem, Chen, Brodie, Lawn, Drummond, Kwan,
  and Ge]{Choong2020}
Jiun Choong, Haris Hakeem, Zhibin Chen, Martin Brodie, Nicholas Lawn, Tom
  Drummond, Patrick Kwan, and Zongyuan Ge.
\newblock {Application of transformers for predicting epilepsy treatment
  response}.
\newblock \emph{medRxiv}, 2020.

\bibitem[He et~al.(2015)He, Zhang, Ren, and Sun]{He2015}
Kaiming He, Xiangyu Zhang, Shaoqing Ren, and Jian Sun.
\newblock Deep residual learning for image recognition.
\newblock \emph{CoRR}, abs/1512.03385, 2015.

\bibitem[Ba et~al.(2016)Ba, Kiros, and Hinton]{ba2016layer}
Jimmy~Lei Ba, Jamie~Ryan Kiros, and Geoffrey~E. Hinton.
\newblock Layer normalization, 2016.

\bibitem[Bertasius et~al.(2021)Bertasius, Wang, and Torresani]{Bertasius2021}
Gedas Bertasius, Heng Wang, and Lorenzo Torresani.
\newblock {Is Space-Time Attention All You Need for Video Understanding?}
\newblock 2021.

\bibitem[Kingma and Ba(2015)]{Kingma2015}
Diederik~P. Kingma and Jimmy~Lei Ba.
\newblock {Adam: A method for stochastic optimization}.
\newblock \emph{3rd International Conference on Learning Representations, ICLR
  2015 - Conference Track Proceedings}, pages 1--15, 2015.

\bibitem[Srivastava et~al.(2014)Srivastava, Hinton, Krizhevsky, Sutskever, and
  Salakhutdinov]{Srivastava2014}
Nitish Srivastava, Geoffrey Hinton, Alex Krizhevsky, Ilya Sutskever, and Ruslan
  Salakhutdinov.
\newblock {Dropout: A Simple Way to Prevent Neural Networks from Overfitting}.
\newblock \emph{Journal of Machine Learning Research}, 2014.

\bibitem[Ioffe and Szegedy(2015)]{Ioffe2015}
Sergey Ioffe and Christian Szegedy.
\newblock {Batch Normalization: Accelerating Deep Network Training by Reducing
  Internal Covariate Shift}.
\newblock feb 2015.

\bibitem[Yu et~al.(2019)Yu, Si, Hu, and Zhang]{Yu2019}
Yong Yu, Xiaosheng Si, Changhua Hu, and Jianxun Zhang.
\newblock {A Review of Recurrent Neural Networks: LSTM Cells and Network
  Architectures}.
\newblock \emph{Neural Computation}, 31\penalty0 (7):\penalty0 1235--1270, jul
  2019.

\bibitem[Brinkmann et~al.(2016)Brinkmann, Wagenaar, Abbot, Adkins, Bosshard,
  Chen, Tieng, He, Muñoz-Almaraz, Botella-Rocamora, Pardo, Zamora-Martinez,
  Hills, Wu, Korshunova, Cukierski, Vite, Patterson, Litt, and
  Worrell]{Benjamin2016}
Benjamin~H. Brinkmann, Joost Wagenaar, Drew Abbot, Phillip Adkins, Simone~C.
  Bosshard, Min Chen, Quang~M. Tieng, Jialune He, F.~J. Muñoz-Almaraz, Paloma
  Botella-Rocamora, Juan Pardo, Francisco Zamora-Martinez, Michael Hills, Wei
  Wu, Iryna Korshunova, Will Cukierski, Charles Vite, Edward~E. Patterson,
  Brian Litt, and Gregory~A. Worrell.
\newblock {Crowdsourcing reproducible seizure forecasting in human and canine
  epilepsy}.
\newblock \emph{Brain}, 139\penalty0 (6):\penalty0 1713--1722, 03 2016.

\bibitem[Antoniades et~al.(2016)Antoniades, Spyrou, Took, and
  Sanei]{Antoniades2016}
Andreas Antoniades, Loukianos Spyrou, Clive~Cheong Took, and Saeid Sanei.
\newblock {Deep learning for epileptic intracranial EEG data}.
\newblock \emph{IEEE International Workshop on Machine Learning for Signal
  Processing, MLSP}, 2016.

\bibitem[Mormann et~al.(2007)Mormann, Andrzejak, Elger, and
  Lehnertz]{Mormann2007}
Florian Mormann, Ralph~G. Andrzejak, Christian~E. Elger, and Klaus Lehnertz.
\newblock {Seizure prediction: The long and winding road}.
\newblock \emph{Brain}, 2007.

\bibitem[Kim and Lee(2017)]{Kim2017}
Soeun Kim and Woojoo Lee.
\newblock {Does McNemar's test compare the sensitivities and specificities of
  two diagnostic tests?}
\newblock \emph{Statistical Methods in Medical Research}, 26\penalty0
  (1):\penalty0 142--154, feb 2017.
\newblock ISSN 0962-2802.
\newblock \doi{10.1177/0962280214541852}.

\bibitem[{Van de Vel} et~al.(2016){Van de Vel}, Cuppens, Bonroy, Milosevic,
  Jansen, {Van Huffel}, Vanrumste, Cras, Lagae, and Ceulemans]{VandeVel2016}
Anouk {Van de Vel}, Kris Cuppens, Bert Bonroy, Milica Milosevic, Katrien
  Jansen, Sabine {Van Huffel}, Bart Vanrumste, Patrick Cras, Lieven Lagae, and
  Berten Ceulemans.
\newblock {Non-EEG seizure detection systems and potential SUDEP prevention:
  State of the art}.
\newblock \emph{Seizure}, 41:\penalty0 141--153, oct 2016.

\end{thebibliography}






\end{document}